\documentclass[aps,prd,nofootinbib,amsmath,amssymb,superscriptaddress,twocolumn,10pt]{revtex4}

\pdfoutput=1

\usepackage{graphicx}
\usepackage{dcolumn}
\usepackage{bm}
\usepackage{amssymb}
\usepackage{latexsym}
\usepackage{booktabs}
\usepackage{amsmath}
\usepackage{multirow}
\usepackage[colorlinks=true, linkcolor=purple, citecolor=blue]{hyperref}

\newcommand{\be}{\begin{equation}}
\newcommand{\ee}{\end{equation}}
\newcommand{\bq}{\begin{eqnarray}}
\newcommand{\eq}{\end{eqnarray}}

\usepackage[usenames,dvipsnames]{xcolor}

\begin{document}

\title{Prospect for constraining holographic dark energy with gravitational wave standard sirens from the Einstein Telescope}

\author{Jing-Fei Zhang}
\affiliation{Department of Physics, College of Sciences, Northeastern
University, Shenyang 110819, China} 
\author{Hong-Yan Dong}
\affiliation{Department of Physics, College of Sciences, Northeastern
University, Shenyang 110819, China} 
\author{Jing-Zhao Qi}
\affiliation{Department of Physics, College of Sciences, Northeastern
University, Shenyang 110819, China} 

\author{Xin Zhang\footnote{Corresponding author}}
\email{zhangxin@mail.neu.edu.cn}
\affiliation{Department of Physics, College of Sciences, Northeastern
University, Shenyang 110819, China} 
\affiliation{Ministry of Education's Key Laboratory of Data Analytics and Optimization
for Smart Industry, Northeastern University, Shenyang 110819, China}
\affiliation{Center for High Energy Physics, Peking University, Beijing 100080, China}

\begin{abstract}
We study the holographic dark energy (HDE) model by using the future gravitational wave (GW) standard siren data observed from the Einstein Telescope (ET) in this work. We simulate 1000 GW standard siren data based on a 10-year observation of the ET to make this analysis. We find that all the cosmological parameters in the HDE model can be tremendously improved by including the GW standard siren data in the cosmological fit. The GW data combined with the current cosmic microwave background anisotropies, baryon acoustic oscillations, and type Ia supernovae data will measure the cosmological parameters $\Omega_{\rm m}$, $H_0$, and $c$ in the HDE model to be at the accuracies of 1.28\%, 0.59\%, and 3.69\%, respectively. A comparison with the cosmological constant model and the constant-$w$ dark energy model shows that, compared to the standard model, the parameter degeneracies will be broken more thoroughly in a dynamical dark energy model. We find that the GW data alone can provide a fairly good measurement for $H_0$, but for other cosmological parameters the GW data alone can only provide rather weak measurements. However, due to the fact that the parameter degeneracies can be broken by the GW data, the standard sirens can play an essential role in improving the parameter estimation.


\end{abstract}
\maketitle

\section{Introduction}\label{sec1}

The discovery of the accelerated expansion of the universe \cite{Riess:1998cb, Perlmutter:1998np} is a milestone in the study of modern cosmology. Dark energy, a dominant component in the universe with a negative pressure, has been proposed to explain the cosmic acceleration \cite{Peebles:2002gy,Copeland:2006wr,Li:2011sd}. But the nature of dark energy up to now still remains mysterious. 

In order to study the nature of dark energy, various theoretical and phenomenological models of dark energy and modified gravity have been proposed. Among these models, the model with a cosmological constant $\Lambda$ and cold dark matter (CDM), also known as the $\Lambda$CDM model, is believed to be the preferred one, because it has only six parameters and can explain various observations quite well \cite{Ade:2013zuv}. However, although the $\Lambda$CDM model is good at fitting the current observational data, it has been always suffering from the severe theoretical puzzles, such as the fine-tuning and coincidence problems \cite{Weinberg:1988cp,Carroll:2000fy}, and thus searching for clues beyond the $\Lambda$CDM model in observation and constructing corresponding cosmological models in theory are an important mission in modern cosmology. 

The simplest extension to the $\Lambda$CDM cosmology is the model with a dark energy having a constant equation-of-state (EoS) parameter $w$, usually known as the $w$CDM model. The apparent shortcoming of this model is that it lacks theoretical roots and such a dark energy is too ad hoc in theory. But, it is still an intriguing phenomenological model in the study of dark energy, since it has only one more parameter than the $\Lambda$CDM model. Therefore, it is important to seek for more realistic dark energy models with more solid theoretical roots. Actually, an interesting example of this kind is provided by the models of holographic dark energy, in which the holographic principle of quantum gravity is combined with the effective quantum field theory \cite{Cohen:1998zx,Li:2004rb}. What is important is that the scenario of holographic dark energy not only can partly resolve the fine-tuning and coincidence problems \cite{Li:2004rb}, but also can fit the current observational data well \cite{Zhang:2005hs,Chang:2005ph,Zhang:2007sh,Li:2009bn,Li:2009jx,Xu:2016grp,Li:2013dha,Wang:2012uf,Wang:2013zca,Zhang:2015rha,He:2016rvp,Cui:2015oda,Zhang:2015uhk,Guo:2015gpa,Feng:2016djj,Wang:2016tsz,Zhang:2017rbg,Zhao:2017urm,Cui:2017idf,Feng:2017mfs,Li:2017usw,Feng:2018yew,Guo:2018ans,Zhang:2014ija}. Currently, the original model of holographic dark energy (HDE) \cite{Li:2004rb} is still a competitive model among the many dark energy models in the aspect of fitting the observations \cite{Xu:2016grp}. Similar to the $w$CDM model, the HDE model also has only one more parameter than the $\Lambda$CDM model. It should also be pointed out that the other two well-known models of this kind, i.e., the new agegraphic dark energy (NADE) model \cite{Wei:2007ty,Zhang:2008mb,Cui:2009ns,Liu:2009xb,Zhang:2009qa,Liu:2010ci,Zhang:2010im,Li:2010ak,Li:2012xm,Zhang:2012pr,Zhang:2013lea} and the holographic Ricci dark energy (RDE) model \cite{Gao:2007ep,Zhang:2009un,Feng:2009hr,Fu:2011ab,Cui:2014sma,Zhang:2014sqa,Yu:2015sla,Cai:2008nk}, have been convincingly excluded by the current observations \cite{Xu:2016grp}.

Currently, the most powerful cosmological probes are provided by the cosmic microwave background (CMB) anisotropies measurements, the baryon acoustic oscillations (BAO) measurements, and the type Ia supernovae (SN) observations. Some important cosmological parameters have been precisely measured by the combination of CMB, BAO, and SN. But, there are still annoying problems in the field of cosmological parameter estimation. For example, several important other parameters beyond the standard $\Lambda$CDM model, such as the EoS of dark energy, the neutrino mass, and the tensor-to-scalar ratio, still cannot be accurately measured \cite{Ade:2013zuv}. In addition, there are still inconsistencies between some observations, and there are degeneracies between some parameters \cite{Ade:2013zuv}. Therefore, we actually need other new cosmological probes other than these traditional optical cosmological probes. In fact, in the future the gravitational wave standard sirens would play an essential role in the cosmological parameter measurement \cite{Zhang:2019ylr}. 

The sources of gravitational waves (GWs) can be used as standard sirens in cosmology, which was first proposed by Schutz \cite{Schutz:1986gp} and subsequently discussed by Holz and Hughes \cite{Holz:2005df}. Actually, the first detection of GWs generated by the binary neutron star (BNS) merger on August 17, 2017 (known as GW170817) \cite{TheLIGOScientific:2017qsa} has initiated the new area of multi-messenger astronomy \cite{GBM:2017lvd}. With the help of the multi-messenger observation of this event, the Hubble constant has been independently determined \cite{Abbott:2017xzu}. The main advantage of this standard siren method is that it avoids using the cosmic distance ladder. The error of this measurement result of the Hubble constant is still large, around 15\%, because only one data point is used. In the future, more low-redshift standard siren data will be accumulated, and thus the error will be decreased to $15\%/\sqrt{N}$, with $N$ being the number of low-redshift standard siren data. Thus, 50 data will lead to a 2\% measurement of the Hubble constant \cite{Chen:2017rfc}. Actually, in the near future, the KAGRA and LIGO-India will join the existing GW detector network, and then the error of the $H_0$ measurement will become smaller, around $13\%/\sqrt{N}$ \cite{Chen:2017rfc}. The third-generation ground-based GW detectors in plan, such as the Cosmic Explorer (CE) and the Einstein Telescope (ET), will have much better detection ability compared to the current advanced LIGO detectors, and so it is expected that the standard sirens would be developed into a powerful cosmological probe.

It is therefore of great interest to know what role the standard sirens would play in the cosmological parameter estimation. Recently, Zhang et al. \cite{Zhang:2018byx} made such an analysis by taking the ET as an example. The ET \cite{ET} is a third-generation ground-based GW detection facility in plan, which has 10 km-long arms and three detectors. Compared to the advanced LIGO, the ET has a much wider detection frequency range and a much better detection sensitivity. Thus, much more BNS merger events in much deeper redshifts can be observed by the ET. By a conservative estimation, in a 10-year run of the ET, about 1000 useful standard sirens can be observed \cite{Zhang:2018byx}. 

It is found in Ref.~\cite{Zhang:2018byx} that the standard sirens are fairly good at measuring the Hubble constant, but for the measurements of other cosmological parameters they are actually not so good. It is shown in Ref.~\cite{Zhang:2018byx} that the measurement of $H_0$ by GW alone is at a 0.3\% precision for the $\Lambda$CDM model, and a 0.5\% precision for the $w$CDM model. The most important finding in Ref.~\cite{Zhang:2018byx} is that the GW standard sirens can be used to break the parameter degeneracies generated by other observations. This is because the standard sirens can measure the absolute luminosity distance. In the $w$CDM model, the contours in the $\Omega_{\rm m}$--$H_0$ plane from the GW data alone and the CMB+BAO+SN data are roughly orthogonal, and thus the degeneracy between the two parameters are thoroughly broken. Furthermore, it is also found that the GW standard sirens cannot provide a good enough measurement for the EoS of dark energy $w$ in the $w$CDM model, with the precision of the $w$ measurement only being about 12\%. As a contrast, an about 4\% measurement for $w$ can be given by the current CMB+BAO+SN observation. However, since the GW standard sirens can break the degeneracy, the combination of CMB+BAO+SN+GW finally can give a 2\% $w$ measurement in the $w$CDM model. Actually, Wang et al. \cite{Wang:2018lun} further showed that the future GW standard sirens observed by the ET can also improve the constraints on the neutrino mass by about 10\%. For other relevant investigations, see, e.g., Refs.~\cite{Cai:2016sby,Cai:2017aea,Cai:2017buj,Cai:2017yww,Sathyaprakash:2009xt,Zhao:2010sz,Li:2013lza,Yang:2017bkv,Feeney:2018mkj,Liao:2017ioi,Wei:2018cov,Du:2018tia,Wei:2019fwp,Fu:2019oll,Yang:2019bpr,Cai:2019cfw,Nunes:2019bjq,Yang:2019vni,Mendonca:2019yfo}. Therefore, it is expected that in the future the GW observation combined with other future optical surveys would be capable of more precisely measuring cosmological parameters and elucidating the nature of dark energy. 

In Ref.~\cite{Zhang:2018byx}, only the $\Lambda$CDM model and the $w$CDM model were considered, and thus we actually need to check for other dynamical dark energy models. It is necessary to check if the standard sirens can play an important role in breaking parameter degeneracies for other dark energy models. In the present work, we will study the models of holographic dark energy with the GW standard sirens from the ET. The aim of this work is to check if the GW standard sirens can also greatly improve the constraints on the HDE model by breaking the parameter degeneracies generated by other observations.

\section{A brief description of the models of holographic dark energy}
\label{sec2} 

The theoretical problem of dark energy is essentially an ultraviolet (UV) problem in the quantum field theory, which is also highly related to theory of gravity, thus the essence of dark energy is a problem of quantum gravity. In the traditional evaluation of the vacuum energy density within the framework of quantum field theory, its value is determined by the sum of the zero-point energy of each mode of all the quantum fields, and thus we have $\rho_{\rm vac}\simeq k_{\rm max}^4/(16\pi^2)$, with $k_{\rm max}$ being the imposed momentum UV cutoff. If the UV cutoff is taken to be the Planck scale (about $10^{19}$ GeV), where the quantum field theory in a classical spacetime metric is expected to break down, the vacuum energy density would exceed the critical density of the universe by some 120 orders of magnitude \cite{Weinberg:1988cp}.

Actually, it has been conjectured that the cosmological constant problem would be thoroughly solved when a full theory of quantum gravity is established. In the present day that the full theory of quantum gravity is still absent, we actually also wish to understand the cosmological constant problem from a point of view of quantum gravity. A typical example of this attempt is the HDE model \cite{Li:2004rb} that considers the gravitational effects and holographic principle in the effective quantum field theory. It is therefore expected that the studies on holographic dark energy might provide significant clues for the bottom-up exploration of a full theory of quantum gravity. 

When the gravity is considered in a quantum field system, the number of degrees of freedom in a given spatial region should be limited owing to the fact that too many degrees of freedom would lead to the formation of a black hole ruining the effectiveness of the quantum field theory \cite{Cohen:1998zx}. Thus, an energy bound is put on the vacuum energy density, $L^{3}\rho_{\rm vac}\leq LM_{\rm pl}^{2}$, where $M_{\rm pl}$ is the reduced Planck mass, which implies that the total energy in a given spatial region with the size $L$ should not exceed the mass of a black hole with the same size \cite{Cohen:1998zx}. Obviously, the infrared (IR) cutoff size of this effective quantum field theory is taken to be the largest length size compatible with is bound. Therefore, a dark energy model based on the effective quantum field theory with a UV/IR duality naturally occurs with the help of the holographic principle. The UV/IR correspondence leads to the fact that the UV problem of dark energy is converted into an IR problem. A given IR scale $L$ saturating that bound will give a dark energy density \cite{Li:2004rb},
\begin{equation}
   \rho_{\rm de}=3c^{2} M_{\rm pl}^{2}L^{-2},
\end{equation}
where $c$ is a dimensionless phenomenological parameter characterizing all of the uncertainties of the theory. It is indicated in this theory that the UV cutoff of the theory is not fixed but runs with the evolution of the IR cutoff, i.e., $k_{\rm max}\propto L^{-1/2}$. Different choices of the IR cutoff $L$ will lead to different holographic dark energy models. 

In this paper, we mainly consider the HDE model \cite{Li:2004rb}. But, as a contrast, we also consider the RDE model \cite{Gao:2007ep}. Although the RDE model has been unfavored by the current observations \cite{Xu:2016grp}, we still consider it in this work since we mainly study what role the GW standard sirens would play in the future parameter estimation and we do not mind if the model is favored by the observations.

\subsection{The HDE model}

The HDE model \cite{Li:2004rb} is defined by choosing the event horizon size of the universe as the IR cutoff. Thus, the dark energy density in the HDE model is given by
\begin{equation}
   \rho_{\rm de}=3c^{2} M_{\rm pl}^{2}R_{\rm eh}^{-2},
\end{equation}
where $R_{\rm eh}$ is the event horizon size defined as
\begin{equation}
R_{\rm eh}(t)=a(t)\int^{\infty}_{t}\frac{dt'}{a(t')}.
\end{equation}

The evolution of the dark energy density in the HDE model is governed by the following differential equations:
\begin{equation}
\begin{aligned}
\frac{1}{E(z)}\frac{dE(z)}{dz}=-\frac{\Omega_{\rm{de}}(z)}{1+z}\left(\frac{1}{2}+\frac{\sqrt{\Omega_{\rm{de}}(z)}}{c}-\frac{3}{2\Omega_{\rm{de}}(z)}\right),\\
\frac{d\Omega_{\rm{de}}(z)}{dz}=-\frac{2\Omega_{\rm{de}}(z)(1-\Omega_{\rm{de}}(z))}{1+z}\left(\frac{1}{2}+\frac{\sqrt{\Omega_{\rm{de}}(z)}}{c}\right),
\end{aligned}
\end{equation}
where $E(z)\equiv H(z)/H_0$ is the dimensionless Hubble parameter. Solving the differential equations (with the initial conditions $E(0)=1$ and $\Omega_{\rm de}(0)=1-\Omega_{\rm m}$) will give the evolutions of both $\Omega_{\rm de}(z)$ and $E(z)$, and then all the cosmological quantities related to the background evolution will be directly derived. The dimensionless parameter $c$ in this model is rather important, since it plays an essential role in determining the properties of dark energy in the HDE model \cite{Zhang:2005yz,Zhang:2006av,Zhang:2006qu,Zhang:2007es,Zhang:2007an,Ma:2007av,Zhang:2009xj}.

\subsection{The RDE model}

The RDE model \cite{Gao:2007ep} is defined by choosing the average radius of the Ricci scalar curvature as the IR cutoff length scale in the theory. The dark energy density in the RDE model can be expressed as
\begin{equation}
   \rho_{\rm de}=3\gamma M_{\rm pl}^{2}(\dot{H}+2H^2),
\end{equation}
where $\gamma$ is a positive constant that is a redefinition in terms of $c$. 

The evolution of the Hubble parameter in this model is determined by the following differential equation:
\begin{equation}
E^2=\Omega_{\rm m} e^{-3x}+\gamma\left({1\over 2}{dE^2\over dx}+2E^2\right),
\end{equation}
where $x=\ln a$. The solution to this differential equation is given by
\begin{equation}
E(z)=\bigg(\frac{2\Omega_{\rm m}}{2-\gamma}(1+z)^{3}+\bigg(1-\frac{2\Omega_{\rm m}}{2-\gamma}\bigg)(1+z)^{(4-\frac{2}{\gamma})}\bigg)^{1/2}.
\end{equation}
The same to the HDE model, the parameter $\gamma$ plays an important role in determining the properties of dark energy in the RDE model \cite{Zhang:2009un}.

\section{Method and data}\label{sec3}

We will simulate the GW standard siren data from the ET and use them to constrain the HDE model and the RDE model. We will investigate whether the standard sirens can tightly constrain the models and whether they can be used to break the parameter degeneracies formed by other observations.  

We first use the current observational data to constrain the models. In this step, we choose the current mainstream mature cosmological probes, i.e., CMB, BAO, and SN. For CMB data, we use the distance priors of the Planck 2018 data~\cite{Aghanim:2018eyx,Chen:2018dbv}. For BAO data, we use the measurements from 6dFGS ($z_{\rm eff}=0.106$) \cite{Beutler:2011hx}, SDSS-MGS ($z_{\rm eff}=0.15$) \cite{Ross:2014qpa}, and BOSS DR12 ($z_{\rm eff}=0.38$, 0.51, and 0.61) \cite{Alam:2016hwk}. For SN data, we use the latest Pantheon compilation \cite{Scolnic:2017caz}. We use the data combination of CMB+BAO+SN to constrain the HDE and RDE models by employing the MCMC package {\tt CosmoMC} \cite{Lewis:2002ah}, and then we take the best-fitted models as the fiducial models to generate the simulated GW standard siren data from the ET. Actually, in this work, we also analyze the cases of the $\Lambda$CDM model and the $w$CDM model, since these two models are taken as reference models in the analysis of the HDE and RDE models.

To simulate the GW standard siren data, we use the simulation method described in Refs.~\cite{Cai:2016sby,Zhao:2010sz,Wang:2018lun,Zhang:2018byx}. So, in this paper we only give a brief description. We simulate 1000 GW standard siren data from the ET, since a conservative estimation tells us that in a 10-year run about 1000 GW standard sirens can be observed by the ET. The most standard siren events are provided by the merger of BNS, and only a small part of them is from the merger of a black hole (BH) and a neutron star (NS). As the same to Refs.~\cite{Cai:2016sby,Wang:2018lun,Zhang:2018byx}, here we also take the ratio between the event numbers of BHNS (the binary system of a BH and a NS) and BNS to be 0.03, according to the prediction of the advanced LIGO-Virgo network. For the mass distributions of NS and BH in the simulation, we randomly sample the mass of NS in the interval $[1,2]~M_{\odot}$ and the mass of BH in the interval $[3,10]~M_{\odot}$, where $M_{\odot}$ is the solar mass, as the same as in Refs.~\cite{Cai:2016sby,Wang:2018lun,Zhang:2018byx}.

The redshift distribution of the GW sources is taken to be of the form \cite{Cai:2016sby,Zhao:2010sz}
\begin{equation}
P(z)\propto \frac{4\pi d_C^2(z)R(z)}{H(z)(1+z)},
\label{equa:pz}
\end{equation}
where $d_C(z)$ is the comoving distance at the redshift $z$ and $R(z)$ denotes the time evolution of the burst rate that takes the form  \cite{Schneider:2000sg,Cutler:2009qv,Cai:2016sby}
\begin{equation}
R(z)=\begin{cases}
1+2z, & z\leq 1, \\
\frac{3}{4}(5-z), & 1<z<5, \\
0, & z\geq 5.
\end{cases}
\label{equa:rz}
\end{equation}
The comoving distance $d_C(z)$ can be calculated by
\begin{equation}
{d_C}(z) = \frac{{1}}{{H_0}}\int_0^z {\frac{{dz'}}{{E(z')}}},
\label{equa:dl}
\end{equation}
where $E(z)=H(z)/H_0$ is given by a cosmological model. Therefore, we can generate a catalog of the GW sources according to the redshift distribution of the GW sources. 

Since the GW amplitude depends on the luminosity distance $d_L$, the information of $d_L$ and $\sigma_{d_L}$ can be obtained from the amplitude of waveform. The strain in the GW interferometers can be written as
\begin{equation}
h(t)=F_+(\theta, \phi, \psi)h_+(t)+F_\times(\theta, \phi, \psi)h_\times(t),
\end{equation}
where the antenna mode functions of the ET (i.e., $F_{+}$ and $F_{\times}$) are \cite{Zhao:2010sz}
 \begin{align}
F_+^{(1)}(\theta, \phi, \psi)=&~~\frac{{\sqrt 3 }}{2}[\frac{1}{2}(1 + {\cos ^2}(\theta ))\cos (2\phi )\cos (2\psi ) \nonumber\\
                              &~~- \cos (\theta )\sin (2\phi )\sin (2\psi )],\nonumber\\
F_\times^{(1)}(\theta, \phi, \psi)=&~~\frac{{\sqrt 3 }}{2}[\frac{1}{2}(1 + {\cos ^2}(\theta ))\cos (2\phi )\sin (2\psi ) \nonumber\\
                              &~~+ \cos (\theta )\sin (2\phi )\cos (2\psi )].
\label{equa:F}
\end{align}
Here ($\theta$,$\phi$) are the angles describing the location of the source relative to the detector, and $\psi$ is the
polarization angle. The three interferometers have $60^\circ$ with each other, so the antenna pattern founctions for the other two interferometers can be easily derived from the above equations. 

The Fourier transform $\mathcal{H}(f)$ of the time domain waveform $h(t)$ is given by
\begin{align}
\mathcal{H}(f)=\mathcal{A}f^{-7/6}\exp[i(2\pi ft_0-\pi/4+2\psi(f/2)-\varphi_{(2.0)})],
\label{equa:hf}
\end{align}
where $\mathcal{A}$ is the Fourier amplitude that is given by
\begin{align}
\mathcal{A}=&~~\frac{1}{d_L}\sqrt{F_+^2(1+\cos^2(\iota))^2+4F_\times^2\cos^2(\iota)}\nonumber\\
            &~~\times \sqrt{5\pi/96}\pi^{-7/6}\mathcal{M}_c^{5/6},
\label{equa:A}
\end{align}
where $d_L=(1+z)d_C$ is the luminosity distance to the source, $\mathcal{M}_c=M \eta^{3/5}$ is the ``chirp mass", $M=m_1+m_2$ is the total mass of coalescing binary with component masses $m_1$ and $m_2$, and $\eta=m_1 m_2/M^2$ is the symmetric mass ratio. Note here that all the masses refer to the observed masses, and the relationship between the observed mass and the intrinsic mass is $M_{\rm obs}=(1+z)M_{\rm int}$. 
$\iota$ is the angle of inclination of the binary's orbital angular momentum with the line of sight. Since the short gamma ray bursts (SGRBs) followed by the mergers are expected to be strongly beamed, the binaries should be orientated nearly face on (i.e., $\iota\simeq 0$) as implied by the coincidence observations of SGRBs. The maximal inclination is about $\iota=20^\circ$. In the simulation, actually averaging the Fisher matrix over the inclination $\iota$ and the polarization $\psi$ with the constraint $\iota<90^\circ$ is roughly the same as taking $\iota=0$ \cite{Li:2013lza}. So, we take $\iota=0$ in the simulation of the GW sources. It is however should be pointed out that in the estimation of the practical uncertainty of the measurement of $d_L$, the impacts of the uncertainty of inclination should be taken into account. In fact, the consideration of the maximal effect of the inclination (between $\iota=0$ and $\iota=90^\circ$) on the signal-to-noise ratio (SNR) leads to a factor of 2. Definitions of other parameters and their values can be found in Ref.~\cite{Cai:2016sby}.

The combined SNR for the network of three independent interferometers can be calculated by
\begin{equation}
\rho=\sqrt{\sum\limits_{i=1}^{3}(\rho^{(i)})^2},
\label{euqa:rho}
\end{equation}
where $\rho^{(i)}=\sqrt{\left\langle \mathcal{H}^{(i)},\mathcal{H}^{(i)}\right\rangle}$, with the inner product defined as \begin{equation}
\left\langle{a,b}\right\rangle=4\int_{f_{\rm lower}}^{f_{\rm upper}}\frac{\tilde a(f)\tilde b^\ast(f)+\tilde a^\ast(f)\tilde b(f)}{2}\frac{df}{S_h(f)},
\label{euqa:product}
\end{equation}
where a tilde above a function denotes the Fourier transform of the function and $S_h(f)$ is the one-side noise power spectral density. In this work, $S_h(f)$ of the ET is taken to be the same as in Ref.~\cite{Zhao:2010sz}.
For the case of the ET, a detection of the GWs is confirmed by using the criterion that the combined SNR is greater than 8 \cite{ET}.

The instrumental error on the measurement of $d_{L}$ can be estimated by using the Fisher information matrix, 
\begin{align}
\sigma_{d_L}^{\rm inst}\simeq \sqrt{\left\langle\frac{\partial \mathcal H}{\partial d_L},\frac{\partial \mathcal H}{\partial d_L}\right\rangle^{-1}}.
\end{align}
It can be easily found that $\sigma_{d_L}^{\rm inst}\simeq d_L/\rho$ due to $\mathcal H \propto d_L^{-1}$. The consideration of the effect from the inclination angle $\iota$ leads to a factor 2 in front of the error, and thus the error is written as
\begin{equation}
\sigma_{d_L}^{\rm inst}\simeq \frac{2d_L}{\rho}.
\label{sigmainst}
\end{equation}
In addition, the error from weak lensing is given by $\sigma_{d_L}^{\rm lens}$ = $0.05z d_L$ \cite{Cai:2016sby}. Therefore, the total error of the measurement of $d_{L}$ can be expressed as
\begin{align}
\sigma_{d_L}&~~=\sqrt{(\sigma_{d_L}^{\rm inst})^2+(\sigma_{d_L}^{\rm lens})^2} \nonumber\\
            &~~=\sqrt{\left(\frac{2d_L}{\rho}\right)^2+(0.05z d_L)^2}.
\label{sigmadl}
\end{align}

We use the method described above to generate the catalogue of the GW standard sirens with their $z$, $d_{L}$, and $\sigma_{d_{L}}$. We simulate 1000 GW standard siren data that are expected to be detected by the ET in its 10-year observation.

For $N$ simulated data points of GW standard sirens, the $\chi^2$ function can be written as
\begin{align}
\chi_{\rm GW}^2=\sum\limits_{i=1}^{N}\left[\frac{\bar{d}_L^i-d_L(\bar{z}_i;\vec{\Omega})}{\bar{\sigma}_{d_L}^i}\right]^2,
\label{equa:chi2}
\end{align}
where $\bar{z}_i$, $\bar{d}_L^i$, and $\bar{\sigma}_{d_L}^i$ are the $i$th redshift, luminosity distance, and error of luminosity distance, and $\vec{\Omega}$ denotes the set of cosmological parameters.

In this work, we will use the data combination of CMB+BAO+SN and the GW data alone to constrain the cosmological models, respectively. From this test, we can observe if the parameter degeneracies formed by the current observation CMB+BAO+SN can be broken by the GW observation. Then, we will further use the data combination CMB+BAO+SN+GW to constrain the models, from which we can learn how the GW standard sirens can help improve the constraints on the cosmological parameters.

Actually, we will investigate four cosmological models, i.e., $\Lambda$CDM, $w$CDM, HDE, and RDE, in this work. The $\Lambda$CDM model is regarded as a standard model of cosmology, and thus it is taken as a reference. The remaining three models, $w$CDM, HDE, and RDE, all have one more parameter than $\Lambda$CDM. The $w$CDM model is actually the simplest extension to the $\Lambda$CDM model. The RDE model is another typical model of the holographic kind scenario. Therefore, in order to comprehensively investigate the HDE model, we wish to make a comparison of it with the other three models.

\begin{figure*}[!htp]
\includegraphics[scale=0.45]{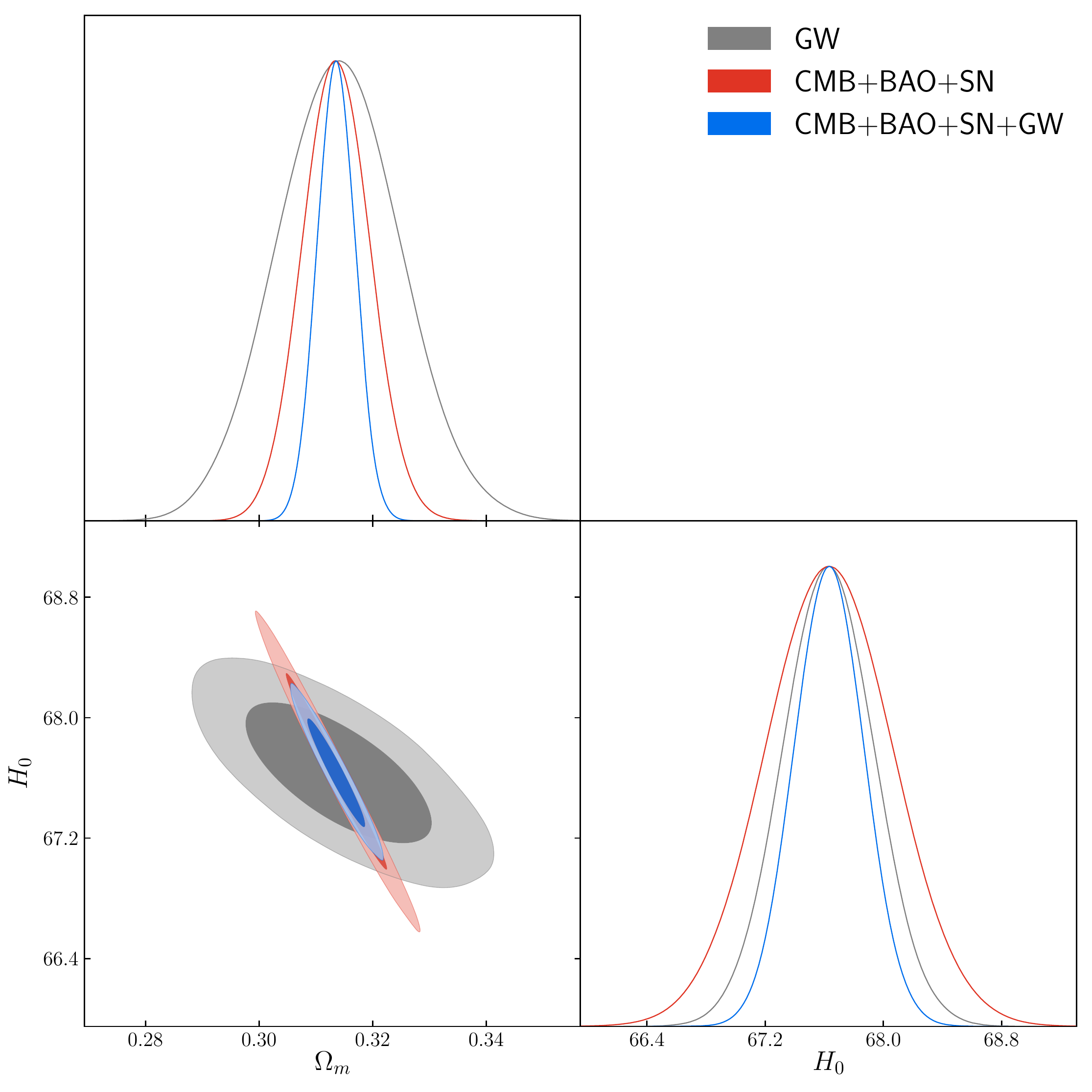}

\centering \caption{\label{fig1} Observational constraints (68.3\% and 95.4\% confidence level) on the $\Lambda$CDM model by using the GW, CMB+BAO+SN, and CMB+BAO+SN+GW data. Here, $H_0$ is in units of km s$^{-1}$ Mpc$^{-1}$.}
\end{figure*}
\begin{figure*}[!htp]
\includegraphics[scale=0.6]{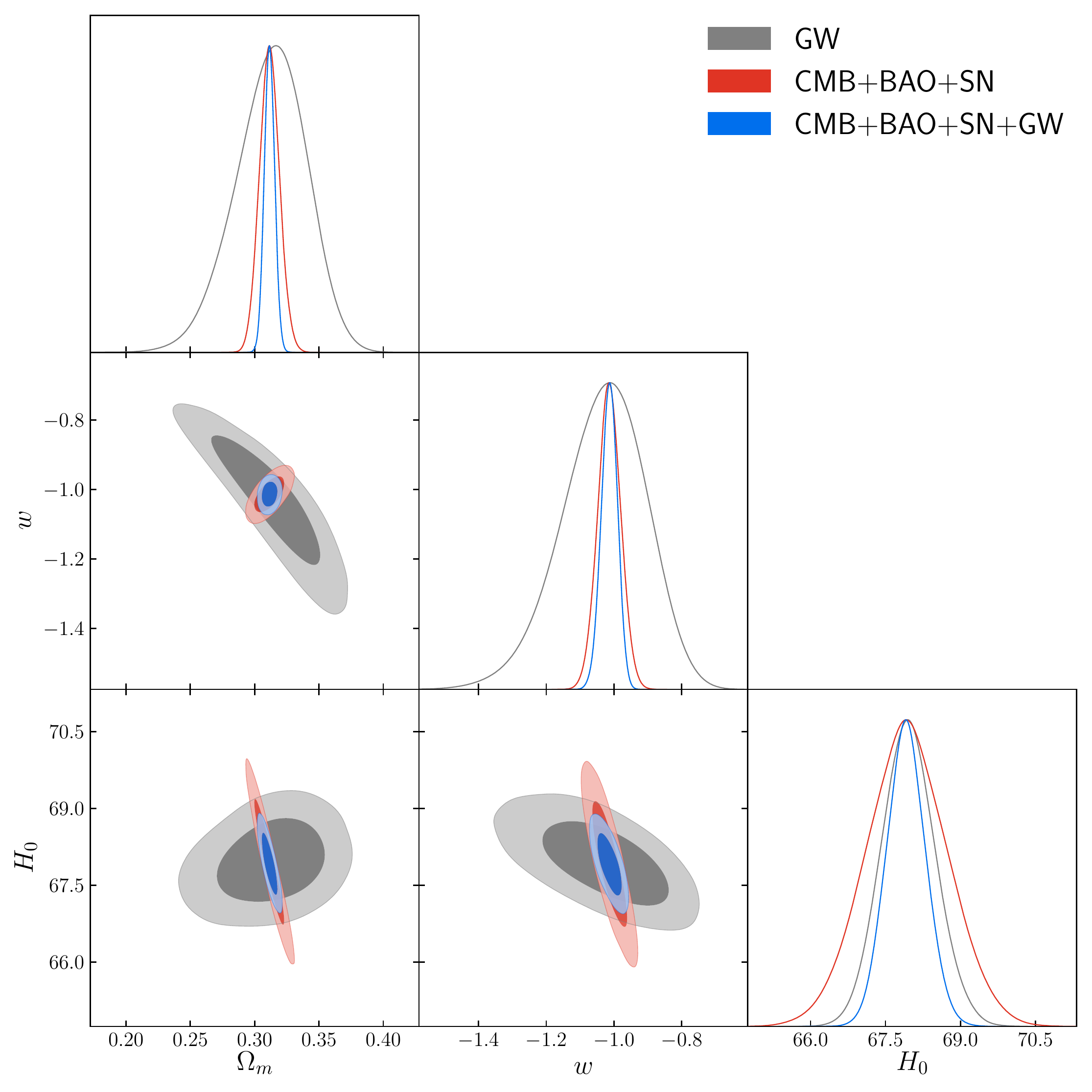}

\centering \caption{\label{fig2} Observational constraints (68.3\% and 95.4\% confidence level) on the $w$CDM model by using the GW, CMB+BAO+SN, and CMB+BAO+SN+GW data. Here, $H_0$ is in units of km s$^{-1}$ Mpc$^{-1}$.}
\end{figure*}
\begin{figure*}[!htp]
\includegraphics[scale=0.6]{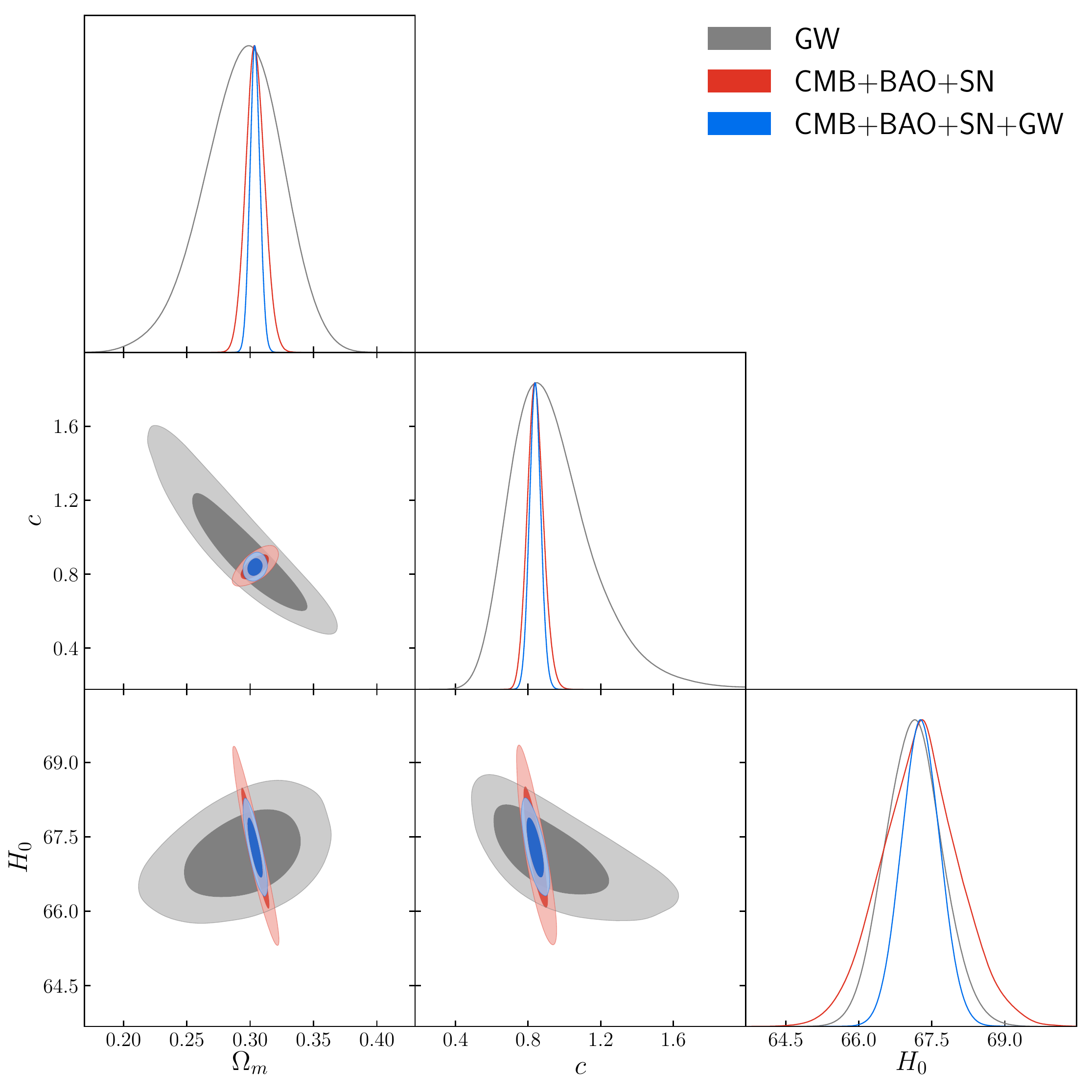}

\centering \caption{\label{fig3} Observational constraints (68.3\% and 95.4\% confidence level) on the HDE model by using the GW, CMB+BAO+SN, and CMB+BAO+SN+GW data. Here, $H_0$ is in units of km s$^{-1}$ Mpc$^{-1}$.}
\end{figure*}
\begin{figure*}[!htp]
\includegraphics[scale=0.6]{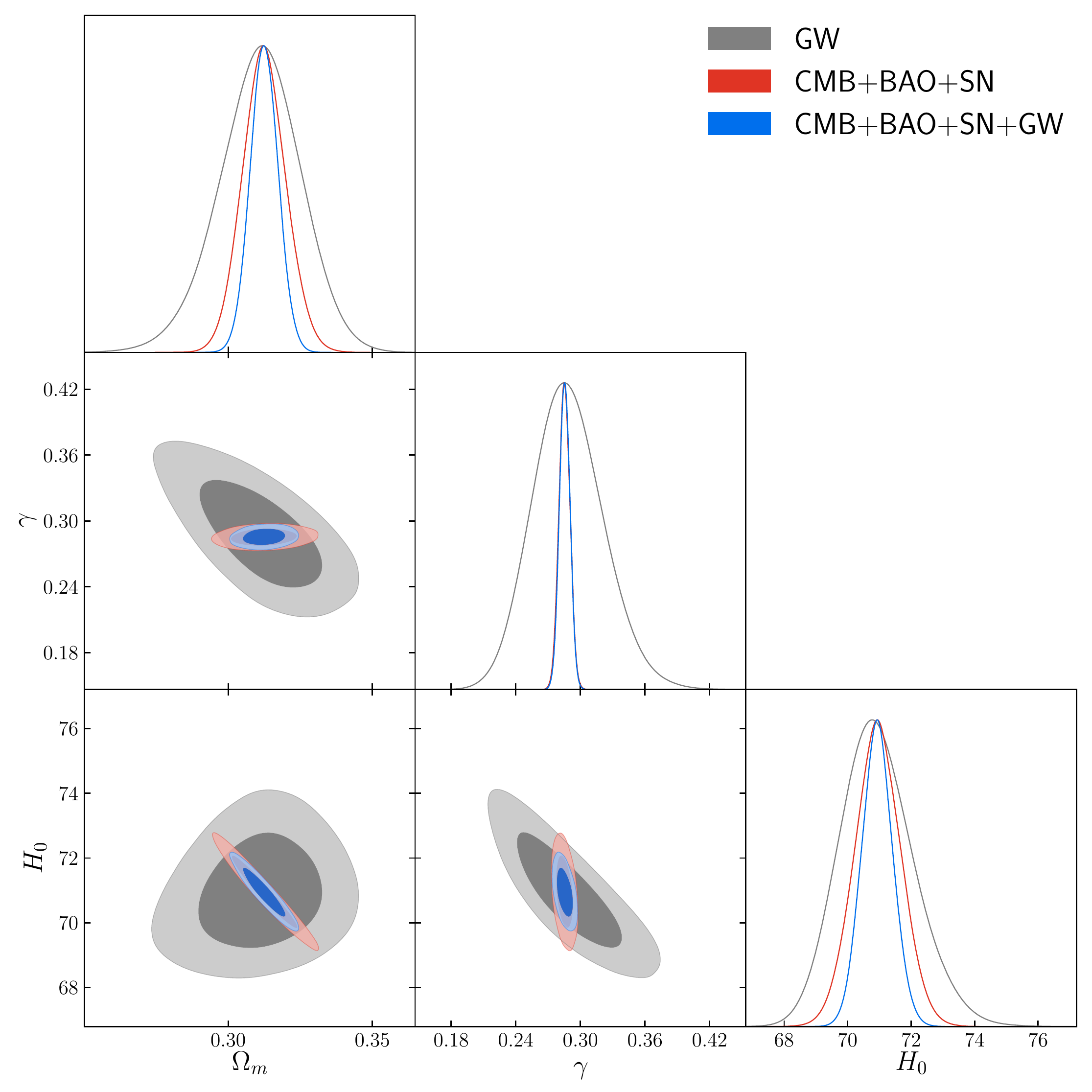}
\centering
\caption{\label{fig4} Observational constraints (68.3\% and 95.4\% confidence level) on the RDE model by using the GW, CMB+BAO+SN, and CMB+BAO+SN+GW data. Here, $H_0$ is in units of km s$^{-1}$ Mpc$^{-1}$.}
\end{figure*}
\begin{table*}[!htp]
        \small
        \centering
        \caption{\label{tab1} Fitting results (68.3\% confidence level) for the $\Lambda$CDM model and the $w$CDM model using GW, CBS, and CBS+GW. Here, CBS stands for CMB+BAO+SN; $H_0$ is in units of km s$^{-1}$ Mpc$^{-1}$.}

		\begin{tabular}{ccccccccccccccccccccccc}

\hline Model&&\multicolumn{3}{c}{$\Lambda$CDM}&&\multicolumn{3}{c}{$w$CDM}&\\
        \cline{1-1}\cline{3-5}\cline{7-9}
 Data&&GW&CBS&CBS+GW&&GW&CBS&CBS+GW&\\
       \hline

 $\Omega_{\rm m}$&&$0.314\pm0.011$&$0.3136\pm0.0059$&$0.3136\pm0.0034$&&$0.312^{+0.030}_{-0.025}$&$0.3116\pm0.0077$&$ 0.3116\pm0.004$\\
 $H_{0}$&&$67.63\pm0.31$&$67.63\pm0.43$&$67.63\pm0.24$&&$67.96\pm0.54$&$67.92\pm0.82$&$67.91\pm0.40$\\
 $w$&&$-$&$-$&$-$&&$-1.03^{+0.14}_{-0.11}$&$-1.014\pm0.034$&$-1.014\pm0.024$\\

\hline
		\end{tabular}
		
\end{table*}
\begin{table*}[!htp]
        \small
        \centering
        \caption{\label{tab2} Fitting results (68.3\% confidence level) for the HDE model and the RDE model using GW, CBS, and CBS+GW. Here, CBS stands for CMB+BAO+SN; $H_0$ is in units of km s$^{-1}$ Mpc$^{-1}$.}

		\begin{tabular}{ccccccccccccccccccccccc}

\hline Model&&\multicolumn{3}{c}{HDE}&&\multicolumn{3}{c}{RDE}&\\
        \cline{1-1}\cline{3-5}\cline{7-9}
 Data&&GW&CBS&CBS+GW&&GW&CBS&CBS+GW&\\
       \hline

 $\Omega_{\rm m}$&&$0.294^{+0.033}_{-0.028}$&$0.304\pm0.0074$&$0.304\pm0.0039$&&$0.311^{+0.015}_{-0.013}$&$0.3125\pm0.0075$&$ 0.3125\pm0.0049$\\
 $H_{0}$&&$67.14\pm0.59$&$67.26\pm0.81$&$67.27\pm0.40$&&$70.9^{+1.1}_{-1.2}$&$70.95\pm0.74$&$70.9\pm0.49$\\
 $c$&&$0.94^{+0.15}_{-0.27}$&$0.841^{+0.041}_{-0.047}$&$0.839\pm0.031$&&$-$&$-$&$-$\\
 $\gamma$&&$-$&$-$&$-$&&$0.289^{+0.029}_{-0.034}$&$0.2852\pm0.0051$&$0.2854\pm0.0049$\\
\hline
		\end{tabular}
		
\end{table*}
\begin{table*}[!htp]
        \small
        \centering
        \caption{\label{tab3} Constraint errors and accuracies for cosmological parameters of the $\Lambda$CDM model and the $w$CDM model using GW, CBS, and CBS+GW. Here, CBS stands for CMB+BAO+SN; $H_0$ is in units of km s$^{-1}$ Mpc$^{-1}$.}

		\begin{tabular}{cccccccccccccccccc}

\hline Model&&\multicolumn{3}{c}{$\Lambda$CDM}&&\multicolumn{3}{c}{$w$CDM}&\\
        \cline{1-1}\cline{3-5}\cline{7-9}
 Data&&GW&CBS&CBS+GW&&GW&CBS&CBS+GW&\\
       \hline

$\sigma(\Omega_{\rm m})$&&$0.0110$&$0.0059$&$0.0034$&&$0.0276$&$0.0077$&$0.0040$\\
$\sigma(H_{0})$&&$0.31$&$0.43$&$0.24$&&$0.54$&$0.82$&$0.40$\\
$\sigma(w)$&&$-$&$-$&$-$&&$0.126$&$0.034$&$0.024$\\
$\varepsilon(\Omega_{\rm m})$&&$0.0350$&$0.0188$&$0.0108$&&$0.0414$&$0.0185$&$0.0134$\\
$\varepsilon(H_{0})$&&$0.0046$&$0.0064$&$0.0035$&&$0.0079$&$0.0120$&$0.0059$\\
$\varepsilon(w)$&&$-$&$-$&$-$&&$0.1223$&$0.0335$&$0.0237$\\
\hline
		\end{tabular}
		
\end{table*}

\begin{table*}[!htp]
        \small
        \centering
        \caption{\label{tab4} Constraint errors and accuracies for cosmological parameters of the HDE model and the RDE model using GW, CBS, and CBS+GW. Here, CBS stands for CMB+BAO+SN; $H_0$ is in units of km s$^{-1}$ Mpc$^{-1}$.}

		\begin{tabular}{cccccccccccccccccc}

\hline Model&&\multicolumn{3}{c}{HDE}&&\multicolumn{3}{c}{RDE}&\\
        \cline{1-1}\cline{3-5}\cline{7-9}
 Data&&GW&CBS&CBS+GW&&GW&CBS&CBS+GW&\\
       \hline

  $\sigma(\Omega_{\rm m})$&&$0.0306$&$0.0074$&$0.0039$&&$0.0140$&$0.0075$&$0.0049$\\
 $\sigma(H_{0})$&&$0.59$&$0.81$&$0.40$&&$1.15$&$0.74$&$0.49$\\
 $\sigma(c)$&&$0.218$&$0.044$&$0.031$&&$-$&$-$&$-$\\
$\sigma(\gamma)$&&$-$&$-$&$-$&&$0.0316$&$0.0051$&$0.0049$\\
$\varepsilon(\Omega_{\rm m})$&&$0.1041$&$0.0243$&$0.0128$&&$0.0450$&$0.0240$&$0.0157$\\
$\varepsilon(H_{0})$&&$0.0088$&$0.0120$&$0.0059$&&$0.0162$&$0.0104$&$0.0069$\\
 $\varepsilon(c)$&&$0.2319$&$0.0523$&$0.0369$&&$-$&$-$&$-$\\
$\varepsilon(\gamma)$&&$-$&$-$&$-$&&$0.1093$&$0.0179$&$0.0172$\\
\hline
		\end{tabular}
		
\end{table*}


\section{Results and discussion}\label{sec4} 

In this section, we report the constraint results and make some relevant discussions. The constraint results are shown in Figs.~\ref{fig1}--\ref{fig4}, and summarized in Tables~\ref{tab1}--\ref{tab4}. In Figs.~\ref{fig1}--\ref{fig4}, the constraint results of the $\Lambda$CDM, $w$CDM, HDE, and RDE models are shown, respectively. The posterior distribution contours (68.3\% and 95.4\% confidence level) and curves from GW alone, CMB+BAO+SN, and CMB+BAO+SN+GW are colored by grey, red, and blue, respectively, in these figures. In Tables \ref{tab1} and \ref{tab2}, the fit values of the cosmological parameters for the models are given. In Tables \ref{tab3} and \ref{tab4}, the constraint errors and accuracies of the cosmological parameters are given. Here, $\sigma(\xi)$ is the absolute error and $\varepsilon(\xi)$ is the relative error, of a parameter $\xi$. In these tables, for convenience, we use the abbreviation ``CBS'' to denote the data combination CMB+BAO+SN.

It should be mentioned that the $\Lambda$CDM model and the $w$CDM model have been investigated using the GW standard sirens in Ref.~\cite{Zhang:2018byx}. In the present paper, since we wish to make an uniform comparison for these models, we redo the analysis for them. We note that there are some little differences between this work and Ref.~\cite{Zhang:2018byx}. First, the actual observational data are somewhat different (for CMB, BAO, and SN). Second, in the simulation of the future GW data we omit a step in this work, namely, the Gaussian random distribution for the simulated data, in order to make the central values of the CMB+BAO+SN data and the GW alone data roughly identical in the parameter plane, which is convenient for the comparison for the constraints from different data. The results of them are similar although slight differences exist, and the conclusion is not changed. 

From these figures and tables, we can easily find that the GW standard sirens can constrain $H_0$ rather tightly (the RDE model is an exception, and we will discuss it in the following), but for the other parameters their constraints are weak. Actually, we directly observe from these figures that the GW standard sirens can play an essential role in breaking the parameter degeneracies in all the cases.

The constraints on the $\Lambda$CDM model are shown in Fig.~\ref{fig1}. We find that the GW data alone can provide a 0.46\% measurement for $H_0$, better than the current CBS constraint with a 0.64\% accuracy. The combined CBS+GW data provide a 0.35\% measurement for $H_0$. We also find that the GW data alone cannot very tightly constrain $\Omega_{\rm m}$, with the constraint accuracy only at 3.50\%, worse than the current CBS constraint with the accuracy 1.88\%. However, due to the fact that the degeneracy is broken by the GW standard sirens, the constraint on $\Omega_{\rm m}$ is improved to be at the 1.08\% level by the combined CBS+GW data.

The constraints on the $w$CDM model are shown in Fig.~\ref{fig2}, from which we can clearly see that the contours from the CBS data and the GW alone data are roughly orthogonal in all the parameter planes (i.e., the $w$--$\Omega_{\rm m}$, $H_0$--$\Omega_{\rm m}$, and $H_0$--$w$ planes), so that the degeneracies are thoroughly broken in this case. For this model, the GW data alone provides a 0.79\% measurement for $H_0$, much better than the current CBS constraint with a 1.20\% accuracy, and the combined CBS+GW data can measure $H_0$ to be at the 0.59\% level. We can see that the GW data alone can only provide a weak constraint on $w$, with the accuracy only at 12.23\%, much worse than the current CBS constraint at the 3.35\% level, but the combined CBS+GW data can improve the result to be at the 2.37\% level thanks to the contribution from the GW standard sirens. 

Now, let us see the constraint results of the HDE model, shown in Fig.~\ref{fig3}. From this figure, we find that the situation of this model is very similar to that of the $w$CDM model. It is clearly seen that the contours from the CBS data and the GW alone data are roughly orthogonal in all the parameter planes (i.e., the $c$--$\Omega_{\rm m}$, $H_0$--$\Omega_{\rm m}$, and $H_0$--$c$ planes), so in this case the degeneracies are also thoroughly broken. For the HDE model, we can see that the GW data alone can provide a 0.88\% measurement for $H_0$, also much better than the current CBS constraint with a 1.20\% accuracy, and the combined CBS+GW data can measure $H_0$ to be at the 0.59\% level. For the measurement of $c$, we find that the constraint capability of the GW alone data is rather weak, which can only provide a 23.19\% measurement, much worse than the current CBS constraint with the accuracy of 5.23\%, and the combined CBS+GW data can improve the result to be at the 3.69\% level owing to the degeneracy being broken. 

Finally, let us discuss the case of the RDE model, with the constraint results shown in Fig.~\ref{fig4}. We find that this case is somewhat different from the above three cases. From Fig.~\ref{fig4}, we can see that in this case the parameter degeneracies formed by the current CBS data are also broken by the future GW standard siren data, but owing to the fact that the constraints from the GW data alone are too weak, the combined CBS+GW data do not improve the constraints as much as the above three cases. We find that for this case the GW data alone can only provide a 1.62\% measurement for $H_0$, slightly worse than the current CBS constraint with a 1.04\% accuracy. The combined CBS+GW data provide a 0.69\% measurement for $H_0$. For the measurement of $\gamma$, we find that the GW alone data can only provide a 10.93\% measurement, much worse than the current CBS constraint with the accuracy of 1.79\%, and the combined CBS+GW data can only slightly improve the result to be at the 1.72\% level. Actually, it should be emphasized that the RDE model has been convincingly ruled out by the current observations, because its $\chi^2$ and information criterion values in the cosmological fit are extremely high compared to other models (in particular, the $\Lambda$CDM model) \cite{Xu:2016grp}. The differences of this model from the other models in this study might originate from this fact. 

{It is well-known that there is a significant tension, about 4.4$\sigma$, for the measurements of the Hubble constant $H_0$, between the Planck result based on the base-$\Lambda$CDM cosmology and the distance-ladder result based on the nearby-universe observation \cite{Riess:2019cxk}. The Planck CMB observation prefers a higher value and the distance ladder gives a lower value for the Hubble constant. Actually, the ``Hubble tension'' is now one of the most important problems in the cosmology today, and it is widely believed as a crisis for cosmology. As discussed in Sec.~\ref{sec1}, the GW standard sirens will serve as a third party to make an arbitration for the Hubble tension. The main advantage of GW standard siren observations is that they provide a pure (luminosity) distance measurement, avoiding the complex astrophysical distance ladder and poorly understood calibration process. In fact, they are calibrated directly by theory. In the near future, the five GW detectors (including KAGRA and LIGO-India) will be capable of reducing the measurement error of $H_0$ to about $13\%/\sqrt{N}$, with $N$ being the number of low-redshift standard siren data \cite{Chen:2017rfc}. The ET, as a third-generation ground-based GW detector, will undoubtedly be a very clear discriminator for the Hubble tension, pointing toward new physics or systematic effects.}

{Here, it is also interesting to note that for the RDE model the central value of $H_0$ in the global fit to the CBS data is around 71 km s$^{-1}$ Mpc$^{-1}$ (see Table~\ref{tab2}), indicating that this model is much better in the alleviation of the Hubble tension than other same-type models. However, although the RDE model is useful in releasing the Hubble tension, it has been convincingly excluded by the current observational data, and thus the interest for this model actually has disappeared \cite{Xu:2016grp,Guo:2018ans}. }

In this study, we show that the future GW standard sirens observed by the ET can greatly help improve the constraints on the HDE model. With the help of the GW standard siren data, the cosmological parameters $\Omega_{\rm m}$, $H_0$, and $c$ in the HDE model would be measured at the accuracies of 1.28\%, 0.59\%, and 3.69\%, respectively. The comparison with the $\Lambda$CDM and $w$CDM models shows that the parameter degeneracies will be broken more thoroughly in a dynamical dark energy model (than the standard model). The GW data alone can provide a very good measurement for $H_0$, but can only provide rather weak constraints on other parameters. However, despite that, the GW standard siren observation would help improve the cosmological parameter estimation to a great extent in a joint constraint. The investigation on the RDE model shows that even though this model has been excluded by the current observations, the parameter estimation for it can also be improved by including the GW standard siren data.

{In the final place, we wish to briefly discuss the issue of the synergies of the GW standard siren observation with other sky surveys (optical, near-infrared, or radio experiments) in the future. In the present work, we only consider the combination of the future GW standard siren observation with the current CBS data to show how the GW standard sirens play the key role of breaking the parameter degeneracies in the optical cosmological observations. Actually, it is the precise measurements for the CMB anisotropies that pushed the studies of the cosmos into the era of ``precision cosmology''. However, due to the fact that the (early and current) CMB observations are the measurements for the early universe, the Planck CMB observation solely cannot provide tight constraints on the late-universe physics (e.g., dark energy parameters, neutrino mass, and so forth). The CMB-alone constraints will always lead to some significant degeneracies between cosmological parameters in the extended cosmological models, and therefore the late-universe (low-redshift) observations (such as BAO, SN, redshift-space distortions, weak lensing, and clusters of galaxies) are needed to be combined with the Planck CMB observation to break the parameter degeneracies inherent in the CMB observation. In the future one to two decades, the next-generation, ``Stage IV'', ground-based CMB experiment (CMB-S4, at South Pole, the high Chilean Atacama plateau, and possibly northern hemisphere sites) \cite{Abazajian:2016yjj}, as well as the Stage IV dark energy experiments (such as DESI \cite{Aghamousa:2016zmz}, LSST \cite{Abell:2009aa}, Euclid \cite{Laureijs:2011gra}, and WFIRST \cite{Spergel:2013tha}; they are optical or near-infrared sky survey projects to measure SN, BAO, redshift-space distortions, and weak lensing, with spectroscopic or imagining methods), will definitely provide a dramatic leap forward in our understanding of the fundamental nature of dark energy and the evolution of the universe. In the future, the GW standard sirens will be developed into a new, powerful cosmological probe, as demonstrated in this paper. Actually, there is another promising cosmological probe provided by the neutral hydrogen 21 cm sky survey \cite{Pritchard:2011xb}. The largest radio telescope in the world, the Square Kilometre Array (SKA) \cite{ska}, is scheduled to be constructed in the near future, which will undoubtedly push the 21 cm cosmology into a new era, and largely promote the progress of cosmology in the forthcoming decades \cite{Bacon:2018dui,Xu:2020uws}. The mid-frequency dish array of the SKA (SKA-MID) to be built in South Africa is designed for studying the evolution of the late universe, especially the nature of dark energy (actually, the full SKA can also be classified into the Stage IV dark energy experiments). Some discussions on the forecasted synergies of these future experiments can be found in, e.g., Refs.~\cite{Bacon:2018dui,Xu:2020uws,Maartens:2015mra,Bull:2015nra,Zhang:2019dyq,Jin:2020hmc}. It can be expected that the future gravitational-wave standard siren observations from ground-based and space-based detectors, combined with the Stage IV CMB experiments and the Stage IV dark energy experiments, as well as the neutral hydrogen 21 cm sky survey from the SKA, would greatly promote the development of cosmology.}

\section{Conclusion}\label{sec5} 

The prospect for constraining the HDE model with GW standard sirens observed from the ET is studied in this work. We find that all the cosmological parameters in the HDE model can be tremendously improved by including the GW data in the cosmological fit. The GW data combined with the current CMB, BAO, and SN data will measure the parameters $\Omega_{\rm m}$, $H_0$, and $c$ in the HDE model to be at the accuracies of 1.28\%, 0.59\%, and 3.69\%, respectively.

Through a comparison with the $\Lambda$CDM and $w$CDM models, we show that, compared to the standard model, the parameter degeneracies will be broken more thoroughly in a dynamical dark energy model. Solely using the GW data can provide a fairly good measurement for $H_0$, but for other parameters the GW data alone can only provide rather weak measurements. Although the constraint capability of the GW data for other parameters is weak, due to the fact that the parameter degeneracies can be broken by the GW data, the standard sirens can play an essential role in improving the parameter estimation. It is also shown that, even though the RDE model has been excluded by the current observations, the GW standard siren data will also help improve the parameter estimation for it.

\begin{acknowledgments}
This work was supported by the National Natural Science Foundation of China (Grants Nos.~11975072, 11875102, 11835009, 11690021, and 11522540) and the National Program for Support of Top-Notch Young Professionals.

\end{acknowledgments}



\end{document}